\documentclass[aps,prd,twocolumn,groupedaddress]{revtex4-1}

\usepackage{amsmath}
\usepackage{xspace}
\usepackage{xcolor}
\usepackage{amssymb}
\usepackage{graphicx}

\newcommand{\dix}[1]{\cdot 10^{#1}}
\newcommand{\dif}[1]{\mathrm{d}#1\:} 
\newcommand{\diftrois}[1]{\mathrm{d}^3#1} 
\newcommand{\deriv}[2]{\frac{\mathrm{d}#1}{\mathrm{d}#2}} 
\newcommand{\derivsec}[2]{\frac{\mathrm{d}^2#1}{\mathrm{d}#2^2}} 
\newcommand{\pderiv}[2]{\frac{\partial #1}{\partial #2}} 
\newcommand{\derivop}[1]{\partial_{#1}}		
\newcommand{\norm}[1]{\ensuremath{\left\Vert #1 \right\Vert}} 
\newcommand{\abs}[1]{\ensuremath{\left\vert #1 \right\vert}} 
\newcommand{\vect}[1]{\ensuremath{\vec{#1}}} 

\newcommand{\gdo}[1]{\bigcirc\left(#1\right)} 

\newcommand{\undemi}{{\frac{1}{2}}}

\begin{document}


\title{Dirac states of an electron in a circular intense magnetic field}


\author{Guillaume Voisin }
\email[]{guillaume.voisin@obspm.fr}
\author{Silvano Bonazzola}
\affiliation{LUTh, Observatoire de Paris, PSL Research University, 5 place Jules Janssen, 92190 Meudon, France}
\author{Fabrice Mottez}
\affiliation{LUTh, Observatoire de Paris, PSL Research University - CNRS,  5 place Jules Janssen, 92190 Meudon, France}

\date{\today}

\begin{abstract}
Neutron-star magnetospheres are structured by very intense magnetic fields extending from 100 to $10^5$ km traveled by very energetic electrons and positrons with Lorentz factors up to $\sim 10^7$. In this context, particles are forced to travel almost along the magnetic field with very small gyro-motion, potentially reaching the quantified regime.

We describe the state of Dirac particles in a locally uniform, constant and curved magnetic field in the approximation that the Larmor radius is very small compared to the radius of curvature of the magnetic field lines. 
 
 We obtain a result that admits the usual relativistic Landau states as a limit of null curvature. We will describe the radiation of these states, that we call quantum curvature or synchro-curvature radiation, in an upcoming paper. 
\end{abstract}


\maketitle

\section{Introduction}

Electron and positron states with very low momentum perpendicular to the magnetic field have been of interest in the field of rotating neutron stars magnetospheres almost since their discovery in 1968 \cite{hewish_observation_1968}. Indeed, the community soon realized that the extremely intense rotating magnetic fields of those magnetospheres, ranging from $\sim 10^{4}$ Teslas at the surface of old millisecond pulsars to $\sim 10^{11}$ Teslas at the surface of some magnetars with a typical $\sim 10^{8}$ Teslas \cite{vigano_magnetic_2015}, could generate extremely large  electric-potential gaps along the open magnetic-field lines ( see e.g. \cite{arons_pulsar_2009} for a review) which in turn accelerate charged particles to energies only limited by radiation reaction. It is believed that these magnetospheres are mostly filled with electron and positrons resulting from a cascade of pair creations : pairs are created by quantum-electrodynamics processes involving gamma rays, and in turn radiate their kinetic energy in gamma-rays that make other pairs. The process of radiation is that of an accelerated charge that inspirals around a curved magnetic field. Because the magnetic field $\vect{B}$ is so intense, the gyro-frequence $\omega =\frac{eB}{m\gamma}$  of an electron of charge $-e$, mass $m$ and Lorentz factor $\gamma$, is so large that the momentum perpendicular to the local field is dissipated to very low values almost instantaneously because of synchrotron radiation reaction. It follows that electrons and positrons are believed to remain mostly very close to the local field line, radiating mostly because of their motion along the curved field line rather than perpendicular to it. Such motion and radiation are described either by the synchro-curvature regime (see e.g. \cite{cheng_general_1996,zhang_quantum_1998,harko_unified_2002,vigano_compact_2014, kelner_synchro-curvature_2015}) or the curvature regime \cite{ruderman_theory_1975}, depending on whether the residual perpendicular motion is taken into account or neglected. 
With basic energetic arguments one then realizes that this can lead the particle to fall down in the quantified regime, both because radiation is efficient and because the energy levels are large in intense magnetic fields. This led the community to study transitions between low-lying Landau levels, see in particular the work of \cite{harding_quantized_1987}. However, Landau levels are defined as the states of an electron in a constant uniform magnetic field and therefore are unable to produce transitions of momentum along the magnetic field, no more that they can explain a curved trajectory of the particle. Additionally, one will notice that the case of curvature radiation corresponds to an unphysical motion : a particle of charge $e$ with a velocity $\vect{v}$ aligned with the local magnetic field $\vect{B}$ cannot undergo the Lorentz force $e\vect{v}\wedge\vect{B}$, and therefore cannot follow the magnetic field line.

Therefore, in this paper our purpose is to generalize the quantum motion of electrons and positrons to the motion in a locally uniform, circular and constant magnetic field, within the assumption that the radius of curvature is large compared to the Larmor radius. To this end we found a very precious help in the previous works about the motion of an electron in a constant uniform magnetic field by \cite{huff_motion_1931}, \cite{johnson_motion_1949} and particularly \cite{sokolovternov}. Basing ourselves on the present paper, we will then be able to derive in an upcoming article the radiation of an electron on its lowest perpendicular levels, which could be called quantum synchro-curvature radiation. 

We shall start by setting up the symmetries of the problem in section \ref{secsym}, before deriving the solutions for a Klein-Gordon particle and more generally for the second order Dirac's equation in section \ref{seckg}. Based on those results we derive the full set of Dirac's Hamiltonian proper states in section \ref{secDirac}. Finally in section \ref{secinterpretation} we propose an interpretation of the obtained states.

\section{Symmetries \label{secsym}}


	We consider a particle along a circular magnetic field line of radius $\rho$ and of axis $\vect{x}$, that we call in the following the main circle. 
	
Further we assume that the characteristic extension $\Delta r$ of the wave function perpendicular to the magnetic field is very small compared to the radius of curvature. This, of course, must be checked a posteriori. In this case one may consider that the magnetic field is locally homogeneous upon an error of $\sim\frac{\Delta r}{\rho}B$. 

	Within these assumptions we have locally three symmetries of the system : one rotation around $\vect{x}$, one rotation around the magnetic-field line and one radial translation from the magnetic-field line. This generates a solid torus around the field line. According to Noether's theorem there will be three corresponding conserved quantities, and so three quantum numbers characterizing the proper states of a particle in such a field, to which one has to add one for the spin symmetry : 
	\begin{itemize}
		\item $s$ that quantizes the orthogonal translation,
		\item $l_\bot$ that quantizes the rotation around the field line,
		\item $l_\parallel$ that quantizes the rotation around $\vect{x}$,
		\item $\zeta$ that accounts for the spin orientation. 
	\end{itemize}
	
	The only difference with the assumptions prevailing in the computation of regular Landau states is that the invariance by translation along the magnetic field is replaced by a rotation around the $\vect{x}$ axis. 
	
\section{Second order and Klein-Gordon solutions\label{seckg}}

 In this paper, except otherwise stated, we always assume summation over repeated indices, latin indices for space components and greek for space-time with a metric of signature $(+---)$.
 
We start from Dirac's Hamiltonian for an electron of charge $-e$ with minimal coupling to a classical magnetic field given by a potential $\vect{A}$ 
\begin{equation}
\label{eqDirachamiltonian1}
\hat{H} =  \alpha^{i}\hat{P_{i}} + \beta mc^2 ,
\end{equation}
where $i = \{x,y,z\} $ and the generalized impulsion is given by 
\begin{equation}
\hat{P_{i}} = -i\hbar\left(\derivop{i} + i\frac{e}{\hbar}A_i\right).
\end{equation}

 Dirac's matrices $\alpha^i$ are given in standard representation (e.g. \cite{berestetskii_quantum_1982}\S 20) b

\begin{equation}
\beta = \left(\begin{tabular}{c c}
1 & 0 \\
0 & 1 
\end{tabular}\right), \alpha^i = \left(
\begin{array}{c c}
	0 & \sigma^i \\
	\sigma^i & 0 
\end{array}\right),
\end{equation}
where the $\sigma^i$ are the Pauli matrices : 
\begin{equation}
	\sigma^x = \left( \begin{matrix}
		0 & 1 \\
		1 & 0
	\end{matrix}\right), 
		\sigma^y = \left( \begin{matrix}
		0 & -i \\
		i & 0
	\end{matrix}\right), 
		\sigma^z = \left( \begin{matrix}
		1 & 0 \\
		0 & -1
	\end{matrix}\right) . 
\end{equation}

\begin{figure}
\begin{center}
\includegraphics[width=0.45 \textwidth]{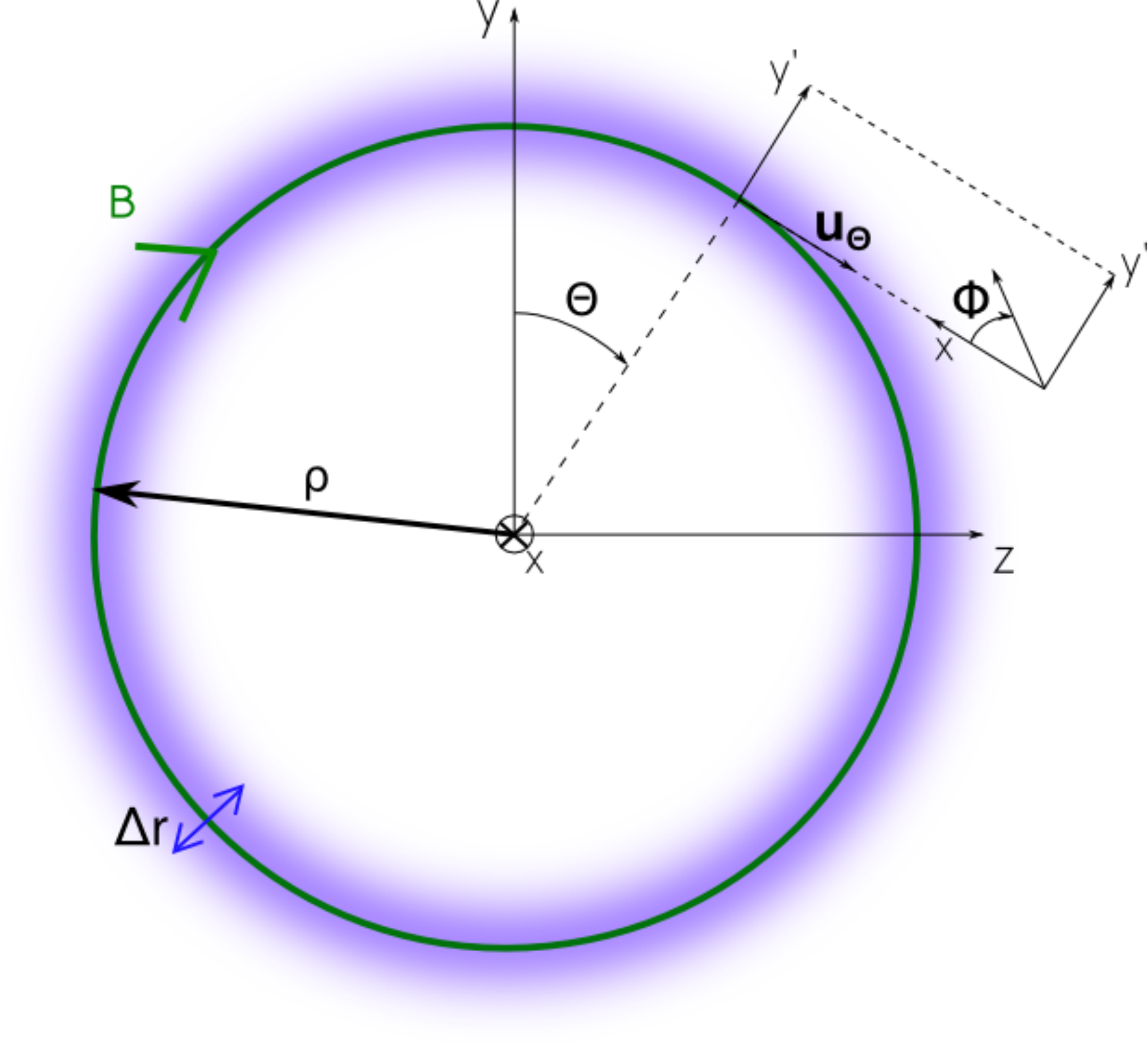}
\caption{\label{figcoord} Representation of a circular magnetic field line (green) of radius $\rho$, that we call in this paper the main circle. The blue shadow around the line represents the wave function of a ground orthogonal level with a characteristic extent $\lambda$. The relation between the toroidal coordinates $(r,\theta,\phi)$ and the cartesian coordinates $(x,y,z)$ is also shown. }
\end{center}
\end{figure}

 We need a coordinate system that makes explicit both the assumed rotation invariance around axis $\vect{x}$ and the part orthogonal to the magnetic field. Such a system is given by the "toroidal" coordinates, represented in figure \ref{figcoord}. Toroidal coordinates are related to the cartesian system $(x,y,z)$ by the homeomorphism
\begin{equation}
 T : (r,\theta,\phi) \rightarrow \left(\begin{array}{l}
 x = r\cos\phi \\ 
 y = \cos\theta (\rho + r\sin\phi) \\ 
 z = \sin\theta(\rho + r\sin\phi)
 \end{array}
  \right),
\end{equation}
where $\theta$ represents the direct angle with respect to the $\vect{y}$ axis in the $(\vect{y}, \vect{z})$ plane.  $\phi$ represents the direct angle with respect to $\vect{x}$ in the plane $(\vect{x}, \vect{y'})$ of the local frame $(\vect{x}, \vect{y'},\vect{u}_\theta)$ image of $(\vect{x}, \vect{y},\vect{z})$ by a rotation of $\theta$ around $\vect{x}$ and $r$ represents the distance to the main circle. In particular we will need 
\begin{equation}
\vect{u}_\theta = \left(0,-\sin\theta,  \cos\theta\right)_{(\vect{x},\vect{y},\vect{z})}.
\end{equation}

In order to write Dirac's equation in this system of coordinates we need to use the Jacobian of $T$, $J_T$, given in by formula \eqref{eqjacobianT} in \ref{aptoroidalcoord}.

Then the covariant quantities, and in particular the impulsion operators transform as follow, 
\begin{equation}
	\hat{p}_{j} = (J_T^{-1})_{ij} \hat{p}_{i'}
\end{equation}
where $i' = \{r, \theta, \phi\}$.

It is can be shown that Dirac's Hamiltonian keeps the same shape as in \eqref{eqDirachamiltonian1} if we express it with Dirac's matrices transformed in a "contravariant" way (see appendix \ref{apcontravariant}), namely
\begin{equation}
	{\alpha}^{i'} = \alpha^j(J_T^{-1})_{i'j}.
\end{equation}

It follows that Dirac's Hamiltonian reads 
\begin{equation}
\hat{H} = \alpha^{i'}\hat{P_{i'}} + \beta mc^2.
\end{equation}

We define a suited expression for the magnetic potential $\vect{A}$ in toroidal coordinates. Since the magnetic field is along $\vec{\derivop{\theta}}$, $A_r$ or $A_\phi$ are its only non zero components. Since we impose rotation invariance $\vect{A}$ does not depend on $\theta$ and the local quasi-uniformity hypothesis implies the dependency in $r$ and $\phi$ should be negligible as long as $r \ll \rho$, and more precisely on the scale of the wave function $\Delta r$. From the expression of the rotational in toroidal coordinates given by \eqref{eqrottoro} in \ref{aptoroidalcoord}, a simple potential yielding a constant magnetic field to lowest order in $r/\rho$ along $\derivop{\theta}$ is 

\begin{equation}
\label{eqmagpotential}
	A^r = 0, A^\theta = 0, A^\phi = -\frac{1}{2}  B .
\end{equation}

Using the metric $g_T$ (\ref{eqmetrictoro}) we obtain the covariant component  \begin{equation}
A_\phi = \frac{1}{2} r^2 B.
\end{equation}
This corresponds to a magnetic field 
\begin{equation}
\vect{B} = B\vect{u}_\theta + \bigcirc\left(r/\rho\right).
\end{equation}

Let's remark that this is compatible with the uniform homogeneous field when $\rho \rightarrow \infty$, as required, since the toroidal coordinates then tend to the cylindrical system.

Following the procedure described in \cite{berestetskii_quantum_1982}, we seek a second order equation of which solutions include Dirac's equation solutions by the taking covariant form of Dirac's equation 
\begin{equation}
\hat{D}\Psi = 0 \Leftrightarrow \left(c\gamma^\mu \hat{P_\mu} - mc^2 \right) \Psi = 0 ,
\end{equation}
and applying to it the operator $\hat{C} = \left( c\gamma^\mu \hat{P_\mu} + mc^2\right)$. One obtains what we call here the second-order Dirac equation 
\begin{equation}
\label{eqsecondorderequation}
\hat{C}\hat{D}\Psi = 0 \Leftrightarrow  \hbar^2 \derivop{t^2}^2\Psi  = \hat{H}_2 \Psi.
\end{equation}

Where $\hat{H}_2$ is the second-order "Hamiltonian" which for the magnetic potential given in \eqref{eqmagpotential} is explicitely given by  
\begin{equation}
\label{eqH2}
\begin{array}{l}
\hat{H}_2 = \\
  (\hbar c)^2 \left[ \derivop{r}^2 + \frac{1}{r}\derivop{r}   + \frac{\cos\phi}{\rho + r\cos\phi} \derivop{r} + \frac{1}{r^2}\derivop{\phi^2}^2 - \frac{\sin\phi}{r(\rho + r\cos\phi)}	\derivop{\phi}  \right.\\
  \left. - 2i\frac{e}{\hbar} A^\phi \derivop{\phi} +  \frac{e^2}{\hbar^2} A^\phi A_\phi - ie \frac{r\sin\phi}{\rho + r\cos\phi} A^\phi \right.\\
  \left. + \frac{1}{(\rho + r\cos\phi)^2}\derivop{\theta^2}^2 - \frac{e}{\hbar}\vect{B}\cdot\vect{\Sigma} - \frac{m^2 c^2}{\hbar^2} \right]
\end{array},
\end{equation}
where $\vect{\Sigma} = \left(\begin{matrix}
\vect{\sigma} & 0 \\
0 & \vect{\sigma}
\end{matrix}\right)$ are the spin $\undemi$ rotation generators in standard representation. The full derivation of \eqref{eqH2} is given in \ref{apH2}. 

Now we notice  that the interaction between the magnetic field and the state of the electron involves the characteristic magnetic length scale
\begin{equation}
\lambda = \left(\frac{2\hbar}{eB}\right)^{1/2}.
\end{equation}

Anticipating on the result, we shall consider that $\lambda$ defines the characteristic perpendicular extent of the wave-function for low-perpendicular-momentum states. This is backed by the fact that the same magnetic length scale plays a similar role in the uniform-magnetic-field case  (see e.g. \cite{sokolovternov}, where $\lambda^{-2}$ is denoted $\gamma$). 
%

We can define a dimensionless coordinate $x \equiv \frac{r}{\lambda}$ and the parameter $\epsilon \equiv \frac{\lambda}{\rho}$. Following our primary assumptions we shall consider that $\epsilon \ll 1$. One can check that this is particularly well verified in the case of a typical pulsar magnetic field of intensity $B \sim  10^8$ Teslas and curvature radius $\rho \sim  10^4$ meters 
\begin{equation}
\epsilon \sim 10^{-16} B_8^{-1/2}\rho_4^{-1},
\end{equation}
where $\rho_4 = \rho/10^4$ and $B_8 = B/10^8$. Notice it could also be true in large particle accelerators, because of the soft (square-root) dependance on the magnetic-field intensity.  

We can now give a quantitative meaning to the assumption of low perpendicular momentum, that is 
\begin{equation}
\label{eqapprox1}
\derivop{x} \sim \derivop{\phi} \ll \epsilon^{-1}.
\end{equation}

The longitudinal momentum can be larger. However we assume that 
\begin{equation}
\label{eqapprox2}
\derivop{\theta} \ll \epsilon^{-3/2},
\end{equation}
and justify this approximation at the end of this section, where \eqref{eqapprox2bis} translates in terms of maximum Lorentz factor the above approximation.

We now rewrite $\hat{H}_2$ \eqref{eqH2} in terms of $x$, $\epsilon $ and $\lambda$ keeping only the lowest order terms in $\epsilon$ 
\begin{equation}
\label{eqH2approx}
\begin{array}{l}
\hat{H}_2 = \\
  \left(\frac{\hbar c}{\lambda}\right)^2 \left[ \derivop{x}^2 + \frac{1}{x}\derivop{x}   + \frac{1}{x^2}\derivop{\phi^2}^2  + 2i \derivop{\phi}   -x^2 \right.\\
  \left. + \epsilon^2 \derivop{\theta^2}^2 - 2\vect{u}_\theta\cdot\vect{\Sigma} - \frac{m^2 c^2 \lambda^2}{\hbar^2} \right] + \gdo{\epsilon}.
\end{array}
\end{equation}

Compared to \eqref{eqH2} the  rightmost term of the two first lines have vanished and we used the fact that $\frac{1}{\rho + r\sin\phi} = \frac{\epsilon}{\lambda} + \gdo{\epsilon^2}$ to simplify the others.
The eigen problem of operator \eqref{eqH2approx} is separable, which would not have been the case if we had considered $ \derivop{\theta} \geq \epsilon^{-3/2}$ since we would have to take into account an additional $2\epsilon^3 x \cos\phi\derivop{\theta^2}^2$. It is also worth noticing that, as expected, this equation is very similar to the one found when solving the uniform field problem in cylindrical coordinates (see \cite{sokolovternov}).

Our symmetry requirements impose that the sought states be proper states not only of the Dirac Hamiltonian but also of two rotation generators 
\begin{eqnarray}
\hat{J}_\theta & = &  -i\hbar \derivop{\phi} + \frac{\hbar}{2} \vect{u}_\theta\cdot \vect{\Sigma}, \\
\hat{J}_x & = & -i\hbar \derivop{\theta} +  \frac{\hbar}{2}\Sigma_x ,
\end{eqnarray}
where $\hat{J}_\theta $ is the angular-momentum operator around the magnetic field centered on the main circle and $J_x$ is the angular-momentum around the axis of the main circle. These operators commute exactly, while commutation with the Dirac Hamiltonian is ensured to order $\epsilon$ ,

\begin{eqnarray}
\left[ \hat{J}_\theta, \hat{J}_x \right] & = & 0, \\
\left[ \hat{H}, \hat{J}_x \right] & = & \gdo{\epsilon}.
\end{eqnarray}

This leads to consider proper states of both operators which are of the form 
\begin{equation}
\begin{array}{l}
\chi_{l_\parallel, j_\bot}\left(\theta,\phi\right) = \frac{1}{2} e^{i j_\bot \phi +i \theta 
  l_\parallel} \\
  \left(\begin{matrix}
 e^{-\frac{i \theta }{2}} \left(b_1 e^{\frac{i \phi }{2}} + b_2 e^{-\frac{i \phi }{2}}\right)- e^{\frac{i \theta
   }{2}} \left(b_1 e^{\frac{i \phi }{2}}-b_2 e^{-\frac{i \phi }{2}}\right) \\
   e^{-\frac{i \theta }{2}} \left(b_1 e^{\frac{i \phi }{2}}+b_2 e^{-\frac{i \phi }{2}}\right)+
   e^{\frac{i \theta }{2}} \left(b_1 e^{\frac{i \phi }{2}} - b_2 e^{-\frac{i \phi }{2}}\right) \\\
   e^{-\frac{i \theta }{2}} \left(b_3 e^{\frac{i \phi }{2}} + b_4 e^{-\frac{i \phi }{2}}\right)-
   e^{\frac{i \theta }{2}} \left(b_3 e^{\frac{i \phi }{2}}-b_4 e^{-\frac{i \phi }{2}}\right) \\
  e^{-\frac{i \theta }{2}} \left(b_3 e^{\frac{i \phi }{2}}+b_4 e^{-\frac{i \phi }{2}}\right) +
   e^{\frac{i \theta }{2}} \left(b_3 e^{\frac{i \phi }{2}} - b_4 e^{-\frac{i \phi }{2}}\right)
\end{matrix}\right)
\end{array},
\end{equation}

where $b_1, b_2, b_3, b_4$ are constants of $\phi$ and $\theta$, and $\hbar j_\bot$ and $\hbar l_\parallel$ are the proper values of $\hat{J}_\theta$ and $\hat{J}_x$ respectively. $l_\parallel$ and $j_\bot$ are half-integers  $...-\frac{3}{2}, -\undemi,  \undemi, \frac{3}{2}...$.


We seek a solution to the eigen problem 
\begin{equation}
\label{eqeigenproblemH2}
\hat{H}_2 \Psi = E^2 \Psi,
\end{equation}
where one can show that $\pm E$ are also a proper energies of the Dirac Hamiltonian $\hat{H}$.
 We seek a solution of the problem \eqref{eqeigenproblemH2} with the help of the ansatz 
\begin{equation}
\label{eqansatz}
	\Psi(x,\theta, \phi) = \chi(\theta, \phi)\Phi(x).
\end{equation}

Inserting this ansatz into \eqref{eqeigenproblemH2} one gets for $\chi$ an eigen problem of the form 
\begin{equation}
\label{eqtheta}
 	\left(\epsilon^2\derivop{\theta^2}^2 - \vect{u}_\theta\cdot\vect{\Sigma}\right)\chi(\theta,\phi) = -C_\theta \chi(\theta),
\end{equation}
where $C_\theta$ is a constant. 
As expected, the operator on the left hand side of equation \eqref{eqtheta} commutes with $\hat{J}_x$ and $\hat{J}_\theta$. 
Therefore $\chi$ is necessarily a combination of $\chi_{l_\parallel,j_\bot}$. Moreover, it is ca be shown that the operator in \eqref{eqtheta} is degenerate with respect to $j_\bot$ but not to $l_\parallel$. Therefore we have that 
\begin{equation}
\chi(\theta,\phi) = \sum_{j_\bot} \chi_{l_\parallel,j_\bot}(\theta,\phi),
\end{equation}
where each $\chi_{l_\parallel,j_\bot}$ has a different set of unknowns $(b_i) = (b_i^{j_\bot})$, $i$ ranging from $1$ to $4$. 

However, we notice that a solution of \eqref{eqtheta} can be found with a combination of only two states, $\chi(\theta,\phi) = \chi_{l_\parallel,j_\bot - 1} + \chi_{l_\parallel,j_\bot} $, if for a given $j_\bot$ one takes $b_1^{j_\bot} = b_3^{j_\bot} = b_2^{j_\bot -1} = b_4^{j_\bot-1 } = 0$, giving 
\newcommand{\lperpm}{l_\bot}
\begin{equation}
\label{eqchi}
\begin{array}{l}
\chi(\theta,\phi) = e^{i l_\parallel \theta} e^{i  \lperpm\phi} \\
\left(\begin{matrix}
e^{-i\frac{\theta}{2}}\left(b_1^{j_\bot - 1} + b_2^{j_\bot}\right) - e^{i\frac{\theta}{2}}\left(b_1^{j_\bot - 1} - b_2^{j_\bot}\right) \\
e^{-i\frac{\theta}{2}}\left(b_1^{j_\bot - 1} + b_2^{j_\bot}\right) + e^{i\frac{\theta}{2}}\left(b_1^{j_\bot - 1} - b_2^{j_\bot}\right) \\
e^{-i\frac{\theta}{2}}\left(b_3^{j_\bot - 1} + b_4^{j_\bot}\right) - e^{i\frac{\theta}{2}}\left(b_3^{j_\bot - 1} - b_4^{j_\bot}\right) \\
e^{-i\frac{\theta}{2}}\left(b_3^{j_\bot - 1} + b_4^{j_\bot}\right) + e^{i\frac{\theta}{2}}\left(b_3^{j_\bot - 1} - b_4^{j_\bot}\right) \\
\end{matrix} \right)
\end{array},
\end{equation}
where we define $ \lperpm \equiv j_\bot - \frac{1}{2}$. With this choice, $\lperpm$ is no longer the exact angular momentum, it will however simplify upcoming calculations and allow a direct comparison with the uniform case as presented in \cite{sokolovternov}. We will give its exact meaning in section \eqref{secinterpretation}.  

Then, $\chi$ is a solution provided that the remaining free coefficients satisfy the following  systems, obtained after inserting \eqref{eqchi} back into \eqref{eqtheta} ,
\begin{equation}
 M_\theta \left(\begin{matrix}
b_{2,4}^{j_\bot} \\
b_{1,3}^{j_\bot - 1}
\end{matrix}\right) = 0,
\end{equation}
where $M_\theta $ is defined by the matrix coefficients
\begin{equation}
\begin{array}{l}
{M_\theta}_{11}  = -\epsilon^2(2 l_\parallel +1 )^2-4 (2-C_\theta), \\
{M_\theta}_{12}  =  (2 l_\parallel  +1 )^2\epsilon^2 - 4 (2 +  C_\theta), \\
{M_\theta}_{21}  = -(1-2 l_\parallel)^2 \epsilon ^2-4 (2- C_\theta) ,\\
{M_\theta}_{22}  =  -(1-2 l_\parallel)^2 \epsilon ^2+4 (C_\theta+2) .\\
\end{array}
\end{equation}

This system has a none trivial solution only if 
\begin{equation}
C_\theta^{\sigma } = \frac{1}{4} \left(4 l_\parallel^2 \epsilon ^2 + \epsilon ^2 + \sigma 4 \sqrt{l_\parallel^2 \epsilon ^4+4}\right),
\end{equation}
which leads to the solution coefficients 
\begin{eqnarray}
\label{eqbcoeffs1}
{b_{2,4}^{j_\bot}}_\sigma & = & c_{1,2}^\sigma \left(2 - l_\parallel\epsilon^2 + \sigma \sqrt{l_\parallel^2 \epsilon ^4+4}\right) ,\\
\label{eqbcoeffs2}
{b_{1,3}^{j_\bot-1}}_\sigma & = & c_{1,2}^\sigma \left(- 2 - l_\parallel\epsilon^2  + \sigma \sqrt{l_\parallel^2 \epsilon ^4+4} \right),
\end{eqnarray}
where $c_1$ and $c_2$ are for now arbitrary constants describing the two proper spaces found for $\left(b_{2}^{j_\bot}, b_{1}^{j_\bot -1 }\right) $ (two first lines of $\chi$) and  $\left(b_{4}^{j_\bot}, b_{3}^{j_\bot -1 }\right) $ (two last lines of $\chi$) respectively. The number $\sigma = \pm 1$ distinguishes two classes of solutions that we shall denote $\uparrow = +1$ and $\downarrow = -1$ for reasons that will become obvious when we see its physical meaning in section \ref{secinterpretation}.


Let's now solve the equation for $\Phi(x)$. After inserting our ansatz \eqref{eqansatz} including the previously found expression for $\chi$, we get the equation 
\begin{equation}
\label{eqradialpart2ndorder}
\left(\derivop{x^2}^2 + \frac{1}{x}\derivop{x} - \frac{\lperpm^2}{x^2}  - x^2 - 2\lperpm \right)\Phi(x)  = -C_x \Phi(x),
\end{equation}
where $C_x$ is a constant to be determined. We give the detailed resolution of this equation in \ref{apresolradialequation}, where we find that 
\begin{eqnarray}
\label{eqphix}
\Phi(x) & = & x^{\lperpm}e^{-\frac{x^2}{2}} L_s^{\lperpm}(x^2) \\
C_x & = & 4 \left( n + \frac{1}{2}\right)  ,
\end{eqnarray}
where $L_s^l$ is a generalized Laguerre polynomial of degree $s$ as defined in \cite{olver_nist_2010} (\S18.5)  by
\begin{equation}
 L_s^l(x) = \sum_{i = 0}^s \frac{(l + i + 1)_{s-i}}{(s-i)! i!}(-x)^i,
\end{equation}
where $(a)_n =\Gamma(a + n)/\Gamma(a) $ is a Pochhammer's symbol. 
  
Besides, $ n = s + \lperpm$ is the primary perpendicular quantum number, $s$ is a positive integer and the perpendicular angular momentum must be positive or null  $\lperpm \geq 0 $ to ensure that the wave function vanishes at infinity and is square-integrable. We will come back later to the interpretation of these quantum numbers. Notice that we use the same notations as in \cite{sokolovternov} for the uniform-magnetic-field case, where the radial dependency has exactly the same form but is expressed with a different coordinate system. Notice as well that \eqref{eqphix} is proportional to a normalization constant that we dropped here for simplicity. Normalization will be determined farther on.


Putting the whole $\Psi$ back into the main equation \eqref{eqeigenproblemH2}, we get the proper energies 
\begin{equation}
 E^2 = m^2c^4 + \frac{\hbar^2 c^2}{\lambda^2} \left(  C_x   + C_\theta^\sigma\right),
\end{equation}
which develops as 
\begin{eqnarray}
E & = & \pm mc^2 \left[ 1 + 2\frac{\hbar\omega_c}{mc^2}\left(n + \frac{1}{2}\right)  + \sigma \frac{\hbar\omega_c}{mc^2}\sqrt{1 + (\epsilon^2 l_\parallel)^2} \right. \nonumber \\
& & \left. + \left(\frac{\hbar\Omega}{mc^2}\right)^2\left(\frac{1}{4} + l_\parallel^2\right)   \right]^\frac{1}{2},
\end{eqnarray}

where $\omega_c = \frac{eB}{m}$ is the cyclotron pulsation and $\Omega = \frac{c}{\rho}$ is the pulsation of the circular trajectory. 

Proper functions $\Psi$ and proper values $E^2$ are the exact solutions of eigen problem of the  approximated operator $\hat{H_2}$ \eqref{eqH2approx}. However, our approximations do not allow us to take meaningfully into account terms of order $\epsilon$ and higher. At this order, the complete solutions of the second order eigen problem \eqref{eqeigenproblemH2} is explicitly given by 

\begin{equation}
\label{eqsolution2nddegre}
\begin{array}{lll}
\Psi_2(x,\theta, \phi) &  = & 4x^{\lperpm}e^{-\frac{x^2}{2}} L_s^{\lperpm}(x^2)e^{il_\parallel\theta}e^{i\lperpm\phi}  \\
 & &  \left(\begin{matrix}
c_1^\sigma \left(\sigma e^{-i\frac{\theta}{2}} + e^{i\frac{\theta}{2}}\right) \\
c_1^\sigma \left(\sigma e^{-i\frac{\theta}{2}} - e^{i\frac{\theta}{2}}\right) \\
c_2^\sigma \left(\sigma e^{-i\frac{\theta}{2}} + e^{i\frac{\theta}{2}}\right) \\
c_2^\sigma \left(\sigma e^{-i\frac{\theta}{2}} - e^{i\frac{\theta}{2}}\right)
\end{matrix}\right) + \gdo{\epsilon} 
\end{array},
\end{equation}
\begin{equation}
\label{eqenergypropre}
E   =  \pm mc^2 \sqrt{ 1 + 2\frac{\hbar\omega_c}{mc^2}\left( n + \frac{1 +\sigma}{2}\right)  + \left(\frac{\hbar\Omega}{mc^2}\right)^2 l_\parallel^2   },
\end{equation}
where in such a development, one has to remind that $l_\parallel$ can be of order $ \epsilon^{-1}$.
In this limit we obtain degenerate states : indeed states with $n , \sigma = +1$ have the same energy as states with numbers $n + 1, \sigma = -1$. The only exception is for what we will from now on call the perpendicular fundamental state : $n = 0, \sigma = -1$, which is non-degenerate.

Before going farther, let's notice that we already obtained the solution of the Klein-Gordon equation for an electron in a circular magnetic field. Indeed, $\hat{H}_2$ corresponds to the Klein-Gordon "Hamiltonian" plus a spin term $\vect{u}_\theta \cdot \Sigma$ . Neglecting this term it comes that $\chi(\theta, \phi) = e^{il_\parallel\theta}e^{i\lperpm\phi}$  and the proper states of energy \eqref{eqenergypropre} are given by
\newcommand{\psiKG}{\Psi_\text{KG}}
\begin{equation}
\label{eqKGsolution}
\psiKG (x, \theta, \phi ) = e^{il_\parallel\theta}e^{i\lperpm\phi}x^{\lperpm}e^{-\frac{x^2}{2}} L_s^{\lperpm}(x^2).
\end{equation}

We now justify a posteriori approximation \eqref{eqapprox2}. Assuming as in  typical pulsar magnetospheres that the motion is dominated by the momentum along the field and that particles are ultra-relativistic with a classical Lorentz factor $\gamma \gg 1$ we obtain, using equation \eqref{eqenergypropre}, 
\begin{equation}
E = \gamma mc^2 = \hbar \Omega l_\parallel + \gdo{\frac{1}{l_\parallel}}.
\end{equation}
This allows to translate approximation \eqref{eqapprox2} in terms of a limit on the Lorentz factor 
\begin{equation}
\label{eqapprox2bis}
\gamma \ll 6\dix{6} \rho_4^{1/2}B_8^{3/4},
\end{equation}
compatible with a variety of pulsar magnetospheres situations. We briefly come back to the interpretation in terms of possible drifts at the end of section \ref{secinterpretation}.

\section{Dirac's equation solutions \label{secDirac}}
\subsection{General solution}
It is can be shown from the derivation of the second order equation \eqref{eqsecondorderequation}  that from any second-order solution a first-order solution can be obtained by applying the $\hat{C}$ operator to it like
\begin{equation}
\Psi(x,\theta, \phi) = \hat{C}e^{-i\frac{E}{\hbar}t}\Psi_2(x,\theta, \phi).
\end{equation}

This is the approach suggested in \cite{berestetskii_quantum_1982}. However, following naively this procedure leads to obtain as many first order solutions as second order solutions while there should be half as many. One can check that we now have 4 independent second-order solutions for each triplet $(n,\lperpm,l_\parallel)$, two for each value of $\sigma$, as shown in equations \eqref{eqbcoeffs1} and \eqref{eqbcoeffs2}.  Moreover, one can check that the obtained second-order solutions are neither directly solutions of $\hat{C}$ or of $\hat{D}$ which implies that proper states for a given energy must be linear combinations of the second order solutions. 

We are going to show that such solutions can be obtained using the combination
\newcommand{\psitest}{\Psi}
\begin{eqnarray}
\label{eqsolutiondirac}
\psitest & = & e^{-i\frac{E}{\hbar}t}\left({\Psi_2}_{\lperpm-1, \sigma = +1} + {\Psi_2}_{\lperpm, \sigma = -1}\right) ,\\
\label{eqenergydirac}
E  & = &  \pm mc^2 \sqrt{ 1 + 2\frac{\hbar\omega_c}{mc^2} n   + \left(\frac{\hbar\Omega}{mc^2}\right)^2 l_\parallel^2   }.
\end{eqnarray}

The state $\psitest$ above is thus defined by the superposition of two states having the same quantum number $s $. The proper energy \eqref{eqenergydirac} can be equivalently defined as ${E^2}_{\lperpm-1, \sigma = +1}$ or ${E^2}_{\lperpm, \sigma = -1}$. We chose the second option in \eqref{eqenergydirac}. Remark that as such $\psitest$ is undefined for $\lperpm = 0$. Prescribing that $ {\Psi_2}_{\lperpm = -1 } = 0$, we find the perpendicular fundamental state as the particular case ${\psitest}_{s = 0, \lperpm = 0}$, 
 
One can show that solving the equation $\hat{D}\psitest = 0$ amounts to solve the linear problem 
\begin{equation}
\label{eqMdsystem}
M_D \left(\begin{matrix}
c_1^- \\
c_2^- \\
c_1^+ \\
c_2^+
\end{matrix}\right) =0,
\end{equation}
Calculations to obtain the matrix $M_D$ are lengthy but appeal to relatively simple operations for which a formal calculation engine can be helpful. We consider the details of it would be of little interest for the reader, for this reason we give here only the main steps. It goes as follow :
\begin{itemize}
\item Divide by the following common factor to isolate the "spinor part" of the equation
\begin{equation}
S \equiv \hat{D}\psitest / (e^{il_\parallel\theta}e^{i(\lperpm-1)\phi}x^{\lperpm}e^{-\frac{x^2}{2}})= 0.
\end{equation}
\item The remaining function can be expanded on the basis of the four orthogonal functions : $\left(e^{a i\frac{\theta}{2}}e^{bi\phi}\right)$, where $a = \pm 1$ and $b=\{0,1\}$. Taking into account the four spinor components, labeled by $j$ hereafter, this gives us 16 coefficients depending on the four unknowns  $(c_1^\downarrow, c_2^\downarrow, c_1^\uparrow , c_2^\uparrow)$ that we call $s_{j,a,b}$. It follows that the equation $\hat{D}\psitest = 0$ reduces to a linear system of 16 equations 
\begin{equation}
\forall j, \forall a, \forall b,  s_{j,a,b} = 0.
\end{equation}
Notice that, at this stage, the coefficients still depend on functions of $x$ . 
\item A lot of these equations are actually equivalent. The coefficients with $a = +1$  are proportional to coefficients with $a=-1$ for any given doublet $(j,b)$. Also notice that the components of the spinor are related two by two : $s_{1,a,b} \propto s_{2,a,b}$ and $s_{3,a,b} \propto s_{4,a,b}$ for all $a$ and $b$. Finally, there are only four a priori independent equations. To fix ideas, we will go on with the system 
\begin{eqnarray}
\label{eqpreMd}\left\{
\begin{array}{lcc}
s_{1,1,1} & = &  0 \\
s_{3,1,1} & = &  0  \\
s_{1,1,0} & = &  0  \\
s_{3,1,0} & = &  0  \\
\end{array}\right. .
\end{eqnarray}
 
\item Using the two relations
\begin{eqnarray}
L_s^{\lperpm + 1}(x^2)  & = & L_{s-1}^{\lperpm + 1}( x^2) + L_s^{\lperpm} (x^2) ,\\
L_s^{\lperpm}(x^2) & = &  \frac{(1 + \lperpm) L_{s}^{\lperpm +1}(x^2)- x^2, L_{s-1}^{\lperpm+2}(x^2)  }{(s + \lperpm + 1)}
\end{eqnarray}
which can be derived from the Laguerre-polynomial recurrence relations given in \cite{olver_nist_2010} \S18.9, one shows that $s_{1,1,1}$ and 
$s_{3,1,1}$ are proportional to $x L_s^{\lperpm+1}$ while $s_{1,1,0}$ and 
$s_{3,1,0}$ are proportional to $L_s^{\lperpm}$. 
\item It follows from the previous point that after dividing each equation by its respective $x$ polynomial as well as $2\hbar c/\lambda$ (to make it dimensionless),  the system \eqref{eqpreMd} gives \eqref{eqMdsystem} with 
\begin{equation}
\begin{array}{l}
M_D = \\
\left(\begin{matrix}
\frac{E - mc^2}{\hbar c / \lambda} &  \epsilon l_\parallel & 0 & 2i \\
 -\epsilon l_\parallel & -\frac{E + mc^2}{\hbar c / \lambda}  & -2i & 0 \\
 0 & -2i(1 + n) & \frac{E - mc^2}{\hbar c / \lambda} &  - \epsilon l_\parallel \\
2i(1 + n ) & 0 &  \epsilon l_\parallel & - \frac{E + mc^2}{\hbar c / \lambda}
\end{matrix} \right)
\end{array}.
\end{equation}

\end{itemize}

The determinant of $M_D$ is null which means, as expected, that the kernel of $M_D$ is not empty. One finds two independent solutions given by
\begin{eqnarray}
\label{eqsoldirac1}
\left(\begin{matrix}
c_1^\downarrow \\
c_2^\downarrow
\end{matrix}\right) = \left(\begin{matrix}
\frac{E + mc^2}{\hbar c / \lambda} \\
-\epsilon l_\parallel
\end{matrix}\right) & \text{ and } 
\left(\begin{matrix}
c_1^\uparrow \\
c_2^\uparrow
\end{matrix} \right) = 
2n\left(\begin{matrix}
0 \\
i
\end{matrix}\right), \\
\label{eqsoldirac2}
\left(\begin{matrix}
c_1^\downarrow \\
c_2^\downarrow
\end{matrix}\right) = \left(\begin{matrix}
-\epsilon l_\parallel \\
\frac{E - mc^2}{\hbar c / \lambda} 
\end{matrix}\right) & \text{ and } 
\left(\begin{matrix}
c_1^\uparrow \\
c_2^\uparrow
\end{matrix} \right) = 
2n\left(\begin{matrix}
i \\
0
\end{matrix}\right).
\end{eqnarray}

Notice that the perpendicular fundamental state comes out naturally from the two solutions \eqref{eqsoldirac1} and \eqref{eqsoldirac2}. Indeed, for $n = 0$ the $\sigma = \uparrow$ coefficients vanish, and the energy becomes $ E = mc^2\sqrt{ 1 + \left(\frac{\hbar\Omega}{mc^2}\right)^2 l_\parallel^2   }$. Then the two solutions are proportional, as expected from the non degeneracy of the perpendicular fundamental, since one finds \eqref{eqsoldirac2} by simply multiplying \eqref{eqsoldirac1} by $-  \frac{E - mc^2}{\hbar c / \lambda} $. The two solutions \eqref{eqsoldirac1} and \eqref{eqsoldirac2} correspond to two spin states that we shall respectively label by $\zeta = -1$ and $\zeta = +1$. Some more details will be given in section \ref{secinterpretation}.

\subsection{Normalization}

We now have obtained the three parts of the wave function. We still need to impose normalization with
\begin{equation}
	\int \diftrois{\vect{x}} \sum_{i=1}^4 \Psi_i^* \Psi_i  = 1.
\end{equation}

We need the Jacobian determinant of the toroidal coordinates 
\begin{equation}
\diftrois{\vect{x}} = \abs{r(\rho + r\sin\phi)} \dif{r}\dif{\theta}\dif{\phi} .
\end{equation}
 Expressing it as a function of the dimensionless variable $x$ ,
 \begin{equation}
\label{eqjactorox}
\diftrois{\vect{x}} = \lambda^2\rho\abs{x(1 + \epsilon x\sin\phi)} \dif{x}\dif{\theta}\dif{\phi} ,
\end{equation}
it becomes obvious that the $\sin\phi$ term can be removed at lowest order in $\epsilon$.

For the integration over $x$ the following integral \cite{olver_nist_2010} 
\begin{equation}
\int_{x=0}^{+\infty} x^{2l} e^{-x^2}\left[L_s^l\left(x^2\right)\right]^2  x \dif{x} = \frac{(s+l)!}{2s!}
\end{equation}
is useful.

%

We get the normalization  
\begin{equation}
\begin{array}{l}
 N = 4\pi \sqrt{\rho \lambda^2 \frac{(n -1)! \left( \left(\frac{E + \zeta mc^2}{\hbar c / \lambda}\right)^2 + (\epsilon l_\parallel)^2 n + 2n^2\right) }{2s!} }
\end{array}.
\end{equation}

\subsection{Complete proper states}

Eventually the proper states of a particle of energy $E$ \eqref{eqenergydirac} in a toroidal magnetic field can be explicitly given by 

\begin{equation}
\label{eqsolutionfinale}
\begin{array}{ll}
\Psi = \frac{1}{N} e^{-\frac{x^2}{2}} e^{i l_\parallel \theta} e^{i \left(\lperpm -\undemi\right) \phi} & \left( e^{-i\frac{\phi}{2}}x^{\lperpm-1}L_s^{\lperpm-1}\left(x^2\right) \chi_\zeta^{\uparrow}(\theta)  \right. \\
&  \left. + e^{i\frac{\phi}{2}} x^{\lperpm}L_s^{\lperpm}\left(x^2\right) \chi_\zeta^{\downarrow}(\theta) \right)
\end{array}.
\end{equation}

The two $\chi^{\sigma}_{\zeta}$ spinors are explicitly given by

\begin{eqnarray}
\label{eqchidown}
\chi_\zeta^{\downarrow} &=& \left(\begin{matrix}
\left(\frac{1+\zeta}{2}\frac{E + mc^2}{\hbar c / \lambda} - \frac{1-\zeta}{2}\epsilon l_\parallel\right) \left(- e^{-i\frac{\theta}{2}} + e^{i\frac{\theta}{2}}\right) \\
\left(\frac{1+\zeta}{2}\frac{E + mc^2}{\hbar c / \lambda} - \frac{1-\zeta}{2}\epsilon l_\parallel\right) \left(- e^{-i\frac{\theta}{2}} - e^{i\frac{\theta}{2}}\right) \\
\left(-\frac{1+\zeta}{2}\epsilon l_\parallel + \frac{1-\zeta}{2}\frac{E - mc^2}{\hbar c / \lambda}\right) \left(- e^{-i\frac{\theta}{2}} + e^{i\frac{\theta}{2}}\right) \\
\left(-\frac{1+\zeta}{2}\epsilon l_\parallel + \frac{1-\zeta}{2}\frac{E - mc^2}{\hbar c / \lambda}\right)  \left(- e^{-i\frac{\theta}{2}} - e^{i\frac{\theta}{2}}\right)
\end{matrix}\right), \\
\label{eqchiup}
\chi_\zeta^{\uparrow} & =  & 2ni\left(\begin{matrix}
\frac{1-\zeta}{2}\left( e^{-i\frac{\theta}{2}} + e^{i\frac{\theta}{2}}\right) \\
\frac{1-\zeta}{2} \left( e^{-i\frac{\theta}{2}} - e^{i\frac{\theta}{2}}\right) \\
\frac{1+\zeta}{2} \left( e^{-i\frac{\theta}{2}} + e^{i\frac{\theta}{2}}\right) \\
\frac{1+\zeta}{2} \left( e^{-i\frac{\theta}{2}} - e^{i\frac{\theta}{2}}\right)
\end{matrix}\right).
\end{eqnarray}

One can see that the constant uniform-magnetic-field case can be recovered by taking the limit $\rho \rightarrow \infty$  in \eqref{eqsolutionfinale} after having performed  the replacements : $\theta \rightarrow z/\rho$, $l_\parallel \rightarrow \rho k^z$ where $\vect{z}$ is the axis along the magnetic field and $k^z$ is the associated wave number.

\section{Interpretation of the quantum numbers}\label{secinterpretation}
In this section we always consider an electron state (positive energy) to simplify the discussion without any loss of generality.

The parallel quantum number $l_\parallel$ quantifies, by construction, the angular momentum around the $\vect{x}$ axis. From the expression of the proper energy we can also interpret $ \hbar\Omega l_\parallel  $ as the "component" of the energy corresponding to the motion along the magnetic field.


We move on to interpreting the perpendicular motion. Our treatment is similar to that of \cite{sokolovternov}. The energy of the motion perpendicular to the magnetic field is quantified by the quantum number $n = s +\lperpm$. It can be interpreted as the quantification of the square of the radius of the trajectory of the electron since the classical gyroradius  can be expressed as $r_g = p_\bot/(m\omega_c)$ with $p_\bot$ the perpendicular momentum and, in case of a purely perpendicular motion, $E^2 = p_\bot^2 + m^2c^4$. We see below that this assertion can be very quickly proven in the classical limit in the particular case $n = \lperpm$.

\begin{figure}
\begin{center}
\includegraphics[width=0.5 \textwidth]{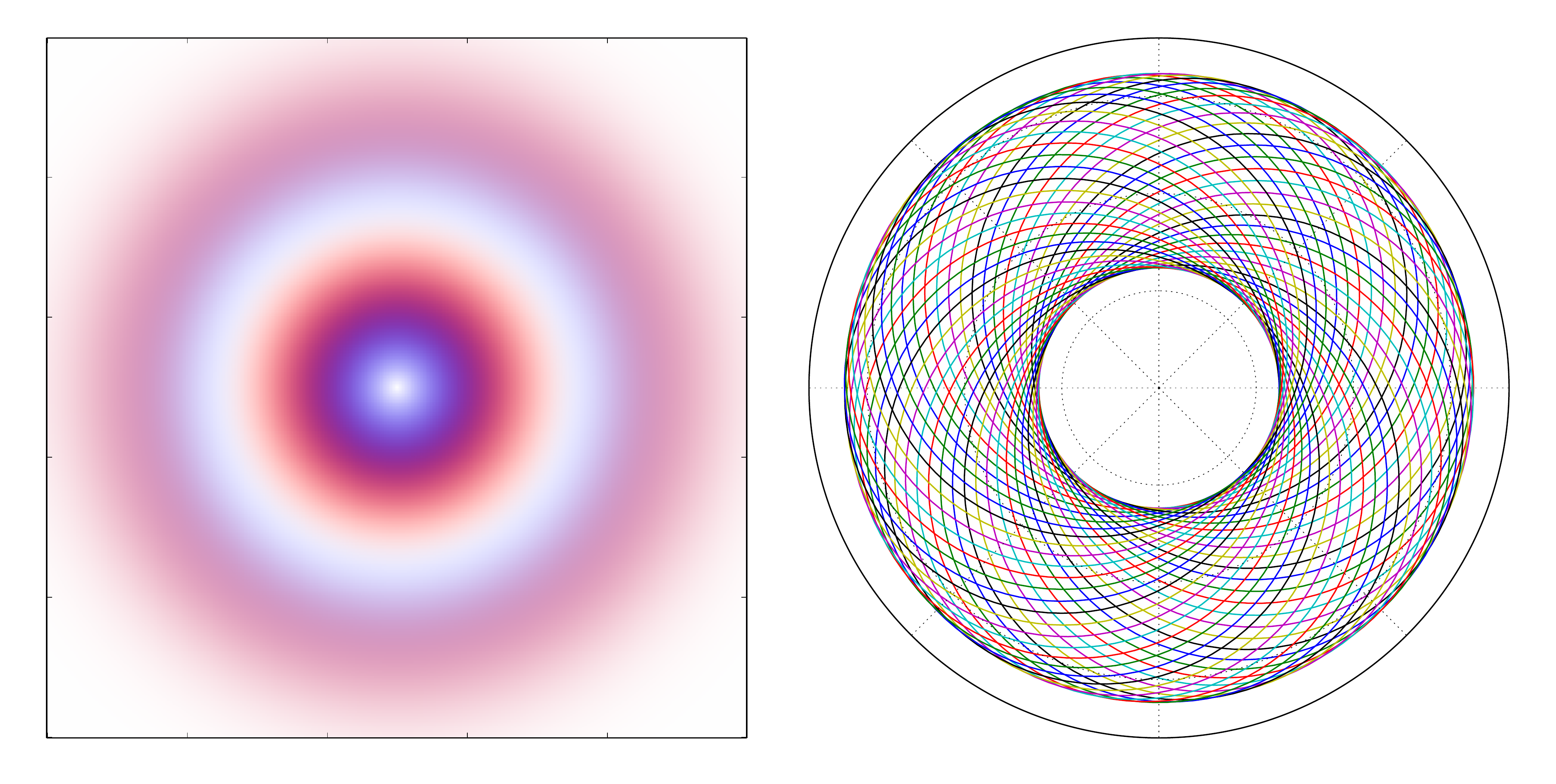}
\caption{\label{figwaveclass} Left panel : probability density of detecting an electron in a state $s=1, \lperpm = 1$ in a plane orthogonal to the magnetic-field line of radius $\rho$. The color goes from blue (inner part of each ring), spin aligned with the magnetic field, to red (outer part of each ring), spin anti-aligned. Here the parallel motion is small, $l_\parallel \lesssim \epsilon^{-1}$ such that both spin components are almost equally important. In the case of a relativistic parallel motion, $l_\parallel > \epsilon^{-1} $, only the anti-aligned (red) component plays a significant role as one can see from equation \eqref{eqsolutionfinale}, \eqref{eqchidown} and \eqref{eqchiup}.
Right panel : representation of a family of off-centered classical trajectories defined by $a = 1.5$ where $a$ is defined in \eqref{eqa}. }
\end{center}
\end{figure}

As we saw in the previous section the wave function is not defined for a strictly negative $\lperpm$. From  a classical point of view this is easily understandable since $\lperpm$ quantifies the angular momentum around the local axis of the magnetic field. Therefore, $\lperpm > 0 $ corresponds to a rotation in the direct sense, which is the orientation that an electron takes under the action of the classical Lorentz force $ \vect{v}\wedge\vect{B}$, where $\vect{v}$ is the speed of the electron. 

Going a little bit deeper, one can show that the solution \eqref{eqsolutionfinale} is a proper state of the angular momentum around the magnetic field $\hat{J}_\theta$ of proper value $\hbar\left(\lperpm - 1/2\right)$. This means that the perpendicular fundamental has a negative angular momentum of $-\hbar/2$. However, it does not mean that the electron classically turns backwards around the magnetic field, but rather that the spin is oriented backward, while the orbital angular momentum is zero. Indeed, one can show that the spinors $\chi_\zeta^\sigma$ are proper states of the operator of projection of the spin onto the main circle of the magnetic field, $\frac{\hbar}{2}\vect{u}_\theta\cdot\vect{\Sigma}$ (the spin part of $\hat{J_\theta}$), with proper values $\hbar\sigma/2$.  We here justify the notation $\uparrow$ or $\downarrow$ for $\sigma =\pm 1$ as meaning that the spin is aligned or anti-aligned with the magnetic field. The perpendicular fundamental state is thus the only purely anti-aligned state, as we will see. Since it has no orbital momentum, one cannot interpret the trajectory of the particle following the magnetic field as the result of the classical Lorentz force but rather as a strictly quantum phenomenon of the interaction between spin and magnetic field.  States with $n > 0$ (i.e. $\lperpm > 0$ or $s > 0$ ) are in a superposed spin state, both aligned and anti-aligned with the magnetic field which results into a degeneracy into two states parametrized by $\zeta$. It is in theory possible to find measurable quantities, hermitian operators that commute with  the Hamiltonian, such that this degeneracy would be lifted and the spin orientation fixed (see e.g. \cite{sokolovternov} or \cite{melrose_quantum_1983} for possibilities in the case of a uniform homogeneous magnetic field). However, it is usually impossible to determine the state of the spin, in particular in astrophysics. We therefore prefer to consider the most general case in which a state of energy $E$ is the superposition of the two $\zeta$ states of \eqref{eqsolutionfinale} combined through a mixing angle $\eta$ ,
\begin{equation}
\Psi = \cos(\eta) \Psi_{\zeta =-1} + \sin(\eta) \Psi_{\zeta =+1}.
\end{equation}
Parametrization by such an angle takes into account the constrain of having a norm of the final state that is still one. Notice that it is impossible to form a purely aligned or anti-aligned state for any value of the mixing angle, as we previously stated. 

We now explain the role of the quantum numbers $s$ and $\lperpm$ and why their role in the energy is degenerated. First consider an electron with $s =0$, then the radial part of the wave function \eqref{eqsolutionfinale} is merely $\propto e^{-x^2}x^{2p}$ where $p = \lperpm-1$ for the anti-aligned term and $p =\lperpm$ for the aligned term. Now, this function is peaked at $x_p = \sqrt{p}$ with an amplitude at the peak of $p^{p}$. This means that, apart for the perpendicular fundamental, the electron always has a double orbit : one  of aligned spin and, a bit further, one of anti-aligned spin, as shown in figure \ref{figwaveclass}. 

Considering a high value of the perpendicular angular momentum one can quickly recover classical results analogous to the uniform magnetic field case. For simplicity we will consider that momentum along the field is zero, $l_\parallel = 0$. From the previous discussion the particle orbits at a distance $r_p \simeq \lambda \sqrt{\lperpm}$ . Expressing $\lperpm$ as a function of the energy one gets 
\begin{equation}
\label{eqlperpclassique}
\lperpm = \frac{E^2 - m^2c^4}{2mc^2\hbar\omega_c}.
\end{equation}
In the classical limit the numerator simply identifies with the square of the perpendicular momentum of the particle $p_\bot^2$. Inserting \eqref{eqlperpclassique} into $r_p$, we obtain the classical (relativistic) Larmor radius :
\begin{equation}
\label{eqlarmorrad}
r_p = r_g = \frac{p_\bot}{eB}
\end{equation}
This is in agreement with the more general result given above. Moreover, it confirms that the typical extent of the wave-function can be taken to be the gyro-radius, at least for high enough quantum numbers, and approximation \eqref{eqapprox1} can we written in a more intuitive way 
\begin{equation}
\label{eqapproxhighqnb}
r_g \ll \rho
\end{equation}

Similarly, one finds that the group velocity of a wave-packet, $v_g = \pderiv{\omega}{k}$, can be found after identifying $\omega = E/\hbar $ and $k = \lperpm / r_p$ : $v_g =  r_p \omega_c/\gamma$ with $\gamma = E/(mc^2)$. We here recognize the classical relativistic gyro-frequency $\omega_c/\gamma$ of an electron in a uniform magnetic field. 

Now, for a same energy we may as well have states of lower $\lperpm$ and higher $s$. This degeneracy also appears, to some extent, in the classical treatment of this problem. Since the radial part of the motion is mostly identical to the uniform-magnetic-field case, we can use the later to better understand the former. We developed in \ref{apclassicalmotion} the Newtonian solution of the uniform problem based on the Hamilton-Jacobi formalism which, because of its parenting with quantum mechanics, allows a formulation in similar terms. In particular, it is found that there are two terms in the perpendicular energy : one related with the angular momentum (noted $p_\theta$ in the classical case) and the other to a shift of distance $r_0$ of the center of the trajectory with respect to the origin of the coordinate. This is summarized in formula \eqref{eqnrjclassique} recalled here,
\begin{equation}
E =  p_\theta\omega_c + \undemi m r_0^2 \omega_c^2.
\end{equation}

We will follow that guide. If $\lperpm$ obviously corresponds to the angular-momentum, $p_\theta$ term, we can show that $s$ corresponds to the second term. The position of the center of the trajectory given by equation \eqref{eqpositioncentrenewt} can be generalized as the operator 
\begin{equation}
\hat{\vect{x_0}} = \hat{\vect{x}} + \frac{\hat{\vect{J}}\wedge\hat{\vect{p}}}{m\omega_c\abs{\hat{\vect{J}}}}
\end{equation}
Where $ \hat{\vect{x}}$ is the position operator, $\hat{\vect{p}}$ the impulsion and  $\hat{\vect{J}}$ the angular momentum with respect to the coordinate origin. For our set of solutions $\hat{\vect{J}} = \hat{J}_\theta $ and the previous operator simplifies to the two components 
\begin{eqnarray}
\hat{x}_0 & = & \hat{x} - \frac{\hat{p}_{y'}}{m\omega_c} ,\\
\hat{y'}_0 & = & \hat{y'} + \frac{\hat{p}_x}{m\omega_c} ,
\end{eqnarray}
where $(x,y')$ are the coordinates locally perpendicular to the magnetic field as defined in  figure \ref{figcoord}. From that, an operator $\hat{r}_0^2 = \hat{x}_0^2 +\hat{y'}_0^2$ is readily obtained. Using the dimensionless coordinate $x = r/\lambda$ 
\begin{equation}
\hat{r}_0^2 = \lambda^2 \left( x^2 -  \derivop{x^2}^2 - \frac{1}{x}\derivop{x} - \frac{1}{x^2}\derivop{\phi^2}^2 + 2i\derivop{\phi} \right),
\end{equation}
where one recognizes the radial part of the second order equation previously solved (see \eqref{eqradialpart2ndorder} or \eqref{eqapradial1} ) except for the $\derivop{\phi}$ term which sign is reversed. From that observation it is straightforward to see that the proper values of $\hat{r}_0^2$ are 
\begin{equation}
 r_0^2 = 4\lambda^2( s  + \undemi).
\end{equation}
The $\undemi$ term comes from the spin interaction that broadens the orbits as we saw previously. Now interpreting the trajectory as an off-centered circle obviously breaks the assumed rotation invariance around the coordinate center. This apparent paradoxe is solved by considering that a proper wave-function is analogous not to a single classical trajectory but to the set of all the trajectories corresponding to the invariants of motion defining the proper state : $n$ or the perpendicular energy and $\lperpm$ or the perpendicular angular momentum. We see from the expression of the trajectory \eqref{eqsolution} that this set is classically parametrized by the constant of integration $\theta_0$. This constant is defined by the initial conditions of the motion, and sets the position of the center of trajectory on the circle of radius $r_0$ centered on the main circle. Then it is obvious that this set is invariant by rotation, as shown in figure \ref{figwaveclass}. Thus, we recover the interpretation of $s$ as characterizing the radial symmetry assumed in section \ref{secsym}.

One notices that in the present solution, the particle remains localized around the magnetic-field line, and therefore there is no drift perpendicular to the line as in the classical theory where the drift is due to the centrifugal force (see e.g. \cite{kelner_synchro-curvature_2015}). This is justified by the fact that we considered only the lowest perpendicular states and a "moderate" longitudinal momentum  \eqref{eqapprox2}  that allows us to  neglect coupling terms between longitudinal motion ($\derivop{\theta}$ terms) and perpendicular motion ($x$, $\derivop{x}$ and $\derivop{\phi}$ terms). We notice that several works on the classical theory of synchro-curvature radiation ( for example \cite{cheng_general_1996,zhang_quantum_1998,harko_unified_2002,vigano_compact_2014})   did not take this drift into account either, and this approximation is widely used for lepton trajectories in pulsar magnetospheres even with Lorentz factors largely above the limit given in \eqref{eqapprox2bis}. Besides, in \cite{kelner_synchro-curvature_2015} the authors show that the effect on radiation of the drift classically results in an effective radius of curvature. 



\section{Conclusion}

In this paper we were able to generalize the relativistic Landau states to the case of circular magnetic field \eqref{eqsolutionfinale}, in the approximation that the curvature radius is large compared to the Larmor radius of the particle while the momentum along the field is not excessively large (\eqref{eqapprox1} or \eqref{eqapproxhighqnb} and \eqref{eqapprox2} ). Our main interest is for applications to the very intense magnetic fields around rotating neutron stars, pulsars and magnetars, in which radiation from very low perpendicular momentum electrons and positrons is believed to be ubiquitous. In an upcoming paper, we will address the problem of radiations from transitions between the states derived in the present paper. We could call them quantum curvature radiation for transitions involving only the ground perpendicular state and more generally quantum synchro-curvature radiation (see e.g. \cite{cheng_general_1996}).

\appendix

\section{Toroidal coordinates toolbox}\label{aptoroidalcoord}

The toroidal coordinates are defined by the following diffeomorphism $T$ 
\begin{equation}
 T : \left(\begin{matrix} r \\ \theta \\ \phi \end{matrix}\right) \rightarrow \left(\begin{matrix} x \\y \\ z \end{matrix} \right) = \left( \begin{matrix} r\cos\phi \\ \cos\theta (\rho + r\sin\phi) \\ \sin\theta(\rho + r\sin\phi) \end{matrix}\right),
\end{equation}
such that surfaces of constant $r$ are torii centered on the circle of radius $\rho > 0$.


The primed quantities denote quantities in the basis  $\left( \pderiv{T}{r},  \pderiv{T}{\theta},  \pderiv{T}{\phi}\right)$.

\subsection{Jacobian}

The Jacobian of this coordinate system is 
\begin{equation}
\label{eqjacobianT}
J_T = \left(
\begin{array}{ccc}
 \cos \phi & 0 & -r \sin \phi  \\
 \cos \theta  \sin\phi  & -\sin \theta  (\rho +r \sin\phi ) & r \cos \theta \cos\phi \\
 \sin \theta \sin\phi  & \cos \theta (\rho +r \sin \phi) & r \sin \theta \cos\phi \\
\end{array}
\right),
\end{equation} 
with determinant  $\det{J_T} = -r (\rho +r \sin\phi)$,
and inverse 
\begin{equation}
J_T^{-1} =  \left(
\begin{array}{ccc}
 \cos\phi & \cos \theta \sin\phi &  \sin\theta \sin\phi \\
 0 & -\frac{\sin \theta}{\rho +r \sin \phi} & \frac{\cos \theta}{\rho +r \sin\phi} \\
 -\frac{\sin\phi}{r} & \frac{\cos\theta \cos\phi}{r} & \frac{\sin\theta \cos \phi}{r} \\
\end{array}
\right).
\end{equation}

\subsection{Transformation of covariant quantities}

Covariant quantities transform like $A_i \rightarrow A'_i$ 
\begin{eqnarray}
		A' = {}^tJ_TA & \Leftrightarrow & A_i' = ({J_T})_{ji}A_j, \\
		A =  {}^t(J_T^{-1}) A' & \Leftrightarrow & A_i = ({J_T^{-1}})_{ji}A'_j.
\end{eqnarray}
Here is an example with the derivation operators , 
\begin{eqnarray}
	 \derivop{x} & = & \cos\phi \derivop{r} - \frac{\sin\phi}{r}\derivop{\phi} ,\\
	 \derivop{y} & = & \cos\theta\sin\phi\derivop{r} - \frac{\sin\theta}{\rho + r\sin\phi}\derivop{\theta} + \frac{\cos\theta\cos\phi}{r}\derivop{\phi} ,\\
	 \derivop{z} & = &  \sin\theta\sin\phi\derivop{r} + \frac{\cos\theta}{\rho + r\sin\phi}\derivop{\theta} + \frac{\sin\theta\cos\phi}{r}\derivop{\phi}.
\end{eqnarray}

The Minkowski metric $\eta = (1,-1,-1,-1)$ transforms according to 
\begin{equation}
M_T = \left(
\begin{matrix}
	1 & 0 \\
	0 & J_T
\end{matrix} \right),
\end{equation}

which gives 
\begin{equation}
\label{eqmetrictoro}
	\eta_T = {}^tM_T g_E M_T = \left(
\begin{array}{cccc}
 1 & 0 & 0 & 0 \\
 0 & -1 & 0 & 0 \\
 0 & 0 & -(\rho +r \sin\phi )^2 & 0 \\
 0 & 0 & 0 & -r^2 \\
\end{array}
\right).
\end{equation}

\subsection{Transformation of contravariant quantities \label{apcontravariant}}
Contravariant quantities transform like $A^i \rightarrow {A'}^i$ 
\begin{eqnarray}
	A'= J_T^{-1} A & \Leftrightarrow & {A'}^i = (J_T^{-1})_{ij}A^{j}, \\
	A = J_T A' & \Leftrightarrow & {A}^i = (J_T)_{ij}{A'}^j.
\end{eqnarray}
This is the case for example of the magnetic potential or of the Dirac matrices $\alpha$ (which are not really contravariant but we use this type of transformation i the text). In particular,

\begin{eqnarray}
	\alpha^r & = & \cos\phi \alpha^x + \cos\theta\sin\phi \alpha^y + 
	\sin\theta\sin \phi \alpha^z, \\
	\alpha^\theta & = & \frac{1}{\rho + r\sin\phi}\left( - \sin\theta\alpha^y + \cos\theta \alpha^z \right) ,\\
	\alpha^\phi & = & -\frac{\sin\phi}{r}\alpha^x + \frac{\cos\theta\cos\phi}{r} \alpha^y + \frac{\sin\theta\cos\phi }{r}\alpha^z .
\end{eqnarray}

\subsection{Transformation of differential operators}
\subsubsection{Laplacian}
The Laplacian is needed for the kinetic part of the second order Dirac equation,
\begin{equation}
\label{eqlaplaciantoro}
\begin{array}{l}
\nabla_T^2 = \frac{1}{r\abs{\rho + r\sin\phi}} \left( \derivop{r}\left(r\abs{\rho + r\sin\phi}\derivop{r}\right) + \right. \\
\left. \derivop{\theta}\left(\frac{r}{\abs{\rho + r\sin\phi}}\derivop{\theta}\right) +  \derivop{\phi}\left(\frac{\abs{\rho + r\sin\phi}}{r} \derivop{\phi}\right) \right)
\end{array}.
\end{equation}

Practically, we always have $\rho + r\sin\phi > 0 $ in this paper. 


\subsubsection{Divergence}
The divergence can be used to derive the second order Dirac equation and is given by

\begin{eqnarray}
\label{eqdivtoro}
	\nabla_T\cdot A' & = & \frac{1}{\abs{r(\rho+r\sin\phi)}} \left( \derivop{r}\left(\abs{r(\rho+r\sin\phi)}{A'}^r \right) + \right. \nonumber \\
& & \left. \derivop{\theta}\left(\abs{r(\rho+r\sin\phi)} {A'}^\theta\right) + \right. \\
& & \left. \derivop{\phi}\left(\abs{r(\rho+r\sin\phi)} {A'}^\phi\right) \right) \nonumber.
\end{eqnarray}

\subsubsection{Rotational of covariant components}
We need the rotational of the magnetic covariant vector which gives the magnetic field
\begin{equation}
B^{x'} =  \left(\nabla_T\wedge ({A'}_i)\right)^{x'},
\end{equation}
which explicitly reads 
\begin{equation}
\label{eqrottoro}
\begin{array}{l}
\begin{array}{l} 
B^{r} = -\left(-\frac{\abs{\rho +r \sin \phi}}{r}\derivop{\phi} + 2 \sin \phi\right) A^\theta + \\
 \frac{r}{\abs{\rho +r \sin \phi} }\derivop{\theta} A^\phi 
 \end{array}, \\
\begin{array}{l} B^{\theta} =  -\frac{1}{\abs{\rho +r \sin \phi}} \left( \frac{1}{r} \derivop{\phi}A^r - (2 + r \derivop{r})A^\phi\right) 
 \end{array}, \\
 \begin{array}{l} B^{\phi} =  -\frac{1}{r\abs{\rho +r \sin \phi}} \left(-\derivop{\theta} A^r +\right. \\
  \left(2 \cos \phi \abs{\rho +r \sin \phi} +  \left. \abs{\rho +r \sin \phi}^2\derivop{r} \right) A^\theta \right) 
\end{array}.
\end{array}
\end{equation}

%
%

\section{Second-order Dirac equation in toroidal coordinates \label{apH2}}
\newcommand{\idmat}{\mathbf{1}}

\textit{Greek indices are used for Minskowski space-time of metric signature $(+---)$ while latin indices are used for the spatial part only. $\eta_{\mu\nu}$ represents the Minkowski metric, $\epsilon^{ijk}$ the fully antisymetric (Ricci) pseudo-tensor,  $\idmat$ represents the identity.
 }

We start with the derivation of the second-order Dirac equation in Cartesian coordinates and then turn it into toroidal coordinates. We take into account the coupling of an electron of charge $-e$ to a classical electromagnetic field defined by a four-potential $(A^\mu) = (\Phi/ c, \vect{A})$ through the covariant derivative defined as 
\newcommand{\derivcov}[1]{\mathrm{D}_#1}

\begin{equation}
	\derivcov{\mu} \Psi = \left( \derivop{\mu} + \frac{i}{\hbar}eA_\mu \right) \Psi.
\end{equation}
 For convenience we use the natural units such that $\hbar = c = 1$. Then the Dirac equation reads 
\begin{equation}
\left( i\gamma^\mu\derivcov{\mu} - mc \right) \Psi = 0,
\end{equation}
on which we apply the "squaring" operator $\left( i\gamma^\mu\derivcov{\mu} + mc \right)$.

The second-order Dirac equation then takes the form 
\begin{equation}
\label{eqsquaredDirac0}
- \left( \left( \gamma^\mu \derivcov{\mu} \right) \left( \gamma^\nu \derivcov{\nu}  \right) + m^2 \right) \Psi = 0.
\end{equation}

 Developing the kinetic part one finds 
\begin{equation}
\label{eqgaDsquared}
\left( \gamma^\mu \derivcov{\mu} \right) \left( \gamma^\nu \derivcov{\nu}  \right) = \left( \derivop{}^\mu \derivop{\mu} + (ie)^2 A^\mu A_\mu \right) + ie \{ \gamma^\mu A_\mu, \gamma^\nu \derivop{\nu} \},
\end{equation}
where
\begin{equation}
\label{eqcomutgaA}
\{ \gamma^\mu A_\mu, \gamma^\nu \derivop{\nu} \}  = \{ \gamma^\mu , \gamma^\nu  \} A_\mu\derivop{\nu} + \gamma^\nu\gamma^\mu\derivop{\nu}\left(A_\mu\right),
\end{equation}
and where 
\begin{eqnarray}
\gamma^\nu\gamma^\mu\derivop{\nu}\left(A_\mu\right) & = &  \frac{1}{2}\left[ \gamma^\mu\gamma^\nu\derivop{\mu}\left(A_\nu\right)  + \gamma^\nu\gamma^\mu\derivop{\nu}\left(A_\mu\right) \right]   \\
& = & \frac{1}{2}\left[ \gamma^\mu\gamma^\nu\left(\derivop{\mu} A_\nu - \derivop{\nu}A_\mu\right)  + 2\eta^{\mu\nu}\derivop{\nu}A_\mu \right] \nonumber.
\end{eqnarray}

Using the identities 
\begin{eqnarray}
\{ \gamma^\mu, \gamma^\nu \} & = & 2\eta^{\mu\nu}\idmat, \\
\gamma^i\gamma^j & = & - \delta_{ij} - i\epsilon_{ijk}\Sigma^k, \\
\gamma^0\gamma^i  & = & \alpha^i,
\end{eqnarray}

and recognizing the electromagnetic field tensor,
\begin{equation}
  F_{\mu\nu} = \derivop{\mu} A_\nu - \derivop{\nu}A_\mu,
\end{equation}
from which we get the contravariant components of the electric and magnetic fields (see e.g. \cite{gourgoulhon_special_2013}) 
\begin{eqnarray}
 E^i & = & \eta^{ij}F_{0j} = -F_{0i}, \\
 B^i & = & \epsilon^{ijk}F_{jk},
\end{eqnarray}
we get 
\begin{equation}
\gamma^\nu\gamma^\mu\derivop{\nu}\left(A_\mu\right)  =  -\vect{\alpha}\cdot \vect{E} - i\Sigma^k \underbrace{\epsilon^{ijk}\derivop{i} A_j}_{\mathrm{curl}(A)^k = B^k} +  2\eta^{\mu\nu}\derivop{\nu}A_\mu.
\end{equation}

The anti-commutator \eqref{eqcomutgaA} then becomes 
\begin{equation}
\label{eqcomutgaA2}
\{ \gamma^\mu A_\mu, \gamma^\nu \derivop{\nu} \}  = 2 (A^\mu \derivop{\mu}  + \undemi \derivop{\mu}A^\mu ) - \vect{\alpha}\cdot \vect{E} - i\vect{B}\cdot\vect{\Sigma}.
\end{equation}

Inserting \eqref{eqcomutgaA2} back into \eqref{eqgaDsquared}, and reorganizing the terms a little we obtain
\begin{equation}
\label{eqgaDsquared2}
\begin{array}{l}
\left( \gamma^\mu \derivcov{\mu} \right) \left( \gamma^\nu \derivcov{\nu} \right) = \\
\underbrace{\left( \derivop{}^\mu \derivop{\mu} + 2i\frac{e}{\hbar} ( A^\mu \derivop{\mu}  + \undemi\derivop{\mu}A^\mu ) -  \frac{e^2}{\hbar^2} A^\mu A_\mu \right) }_{(\derivop{\mu} + i\frac{e}{\hbar} A_\mu)^2} - \\
ie \vect{\alpha}\cdot \vect{E} + e\vect{B}\cdot\vect{\Sigma}
\end{array},
\end{equation}
from what we get the same expression for the second-order Dirac equation as in \cite{berestetskii_quantum_1982} (One will pay attention that in Cartesian coordinates $A_x = -A^x$ and that the usual magnetic potential $\vect{A}$ is defined as a contravariant quantity. With the metric signature used here : $A^0 = \Phi/c = A_0$.),
\begin{equation}
\begin{array}{l}
(\hbar c)^2 \left[ \left( \frac{1}{c} \derivop{t} + i\frac{e}{\hbar c}\Phi \right)^2 - \sum_{x \in \{x,y,z\} }  \left( \derivop{x}  + i\frac{e}{\hbar} A_x \right)^2   - \right.\\
\left. i\frac{e}{\hbar c} \vect{\alpha}\cdot \vect{E} + \frac{e}{\hbar}\vect{B}\cdot\vect{\Sigma} + \frac{m^2 c^2}{\hbar^2} \right] \Psi = 0
\end{array}
\end{equation}

We now switch to another spatial coordinate system denoted by primes, with the only assumption that this system is orthogonal. The Jacobian of the transformation is given by $(J^{ij})$. 
In \eqref{eqgaDsquared2} we separate the time components from the space components,
\begin{equation}
\label{eqgaDsquared2}
\begin{array}{l}
\left( \gamma^\mu \derivcov{\mu} \right) \left( \gamma^\nu \derivcov{\nu} \right) = \left(\derivop{t}  + ie \Phi \right)^2 - \\
\left(\Delta' - 2i\frac{e}{\hbar}\left(   {A'}^i\derivop{i}' + \undemi\nabla' \cdot \vect{A'}  \right) + \frac{e^2}{\hbar^2}{A'}^i A'_i\right)  - \\
ie \vect{\alpha}\cdot \vect{E} + e\vect{B}\cdot\vect{\Sigma}
\end{array}.
\end{equation}

Where $ A'_i = J^{ij} A_i $ and $\Delta', \nabla'\cdot$ represent the Laplacian and the divergence in the primed system of coordinates.  We have used the orthogonality of $J_{ij}$ to eliminate cross terms. If only spatial coordinates change the electric field $\vect{E}$ here transforms like a covariant vector and $\vect{\alpha}$ like a contravariant quantity (as shown in the text) such that $\vect{\alpha}\cdot \vect{E}  = \vect{\alpha}'\cdot \vect{E}'$. The rules of transformation of the magnetic field are less straighforward and it might be simpler to just express it as a function of the primed variables without changing its basis. That is the choice of this paper.

In the case proposed in this paper, we use the toroidal coordinates defined in \ref{aptoroidalcoord}, with a Laplacian and a divergence respectively given by \eqref{eqlaplaciantoro} and \eqref{eqdivtoro}. The magnetic potential is assumed to be only along the third direction : $\vect{A'} = (0,0,A^\phi)$.  All replacements made we obtain

\begin{equation}
\begin{array}{l}
 - (\hbar c)^2 \left[  - \frac{1}{c^2} \derivop{t^2}^2  + \right. \\
 \derivop{r}^2 + \frac{1}{r}\derivop{r} + \frac{1}{(\rho + r\cos\phi)^2}\derivop{\theta^2}^2 + \frac{1}{r^2}\derivop{\phi^2}^2 -  \\
2i\frac{e}{\hbar} A^\phi \derivop{\phi} +  \frac{e^2}{\hbar^2} A^\phi A_\phi - \frac{e}{\hbar}\vect{B}\cdot\vect{\Sigma} - \frac{m^2 c^2}{\hbar^2} + \\
\left. \frac{\cos\phi}{\rho + r\cos\phi} \derivop{r} - \frac{\sin\phi}{r(\rho + r\cos\phi)}	\derivop{\phi} -   ie \frac{r\sin\phi}{\rho + r\cos\phi} A^\phi\right] \Psi = 0
\end{array}.
\end{equation}

The non-negligible (see the text) Laplacian terms are on the second line. The terms involving the  magnetic potential are on the third line. All the terms on the fourth line are neglected in this paper, the two leftmost terms coming from the Laplacian and the rightmost term being the divergence.

\section{Resolution of the radial differential equation \label{apresolradialequation}}
In this appendix, we develop the detailed solution of the differential equation \eqref{eqradialpart2ndorder} giving the radial dependency of the proper states of Dirac's equation. Here we recall the equation 
\begin{equation}
\label{eqapradial1}
\left( \derivop{x^2}^2 + \frac{1}{x}\derivop{x} + \frac{1}{x^2}\derivop{\phi^2}^2 + 2i\derivop{\phi} - x^2 + C\right) f(x,\phi) = 0, 
\end{equation}
Where $-C$ is the proper value of the equation, to be determined. 

Assuming the following form for $f$ :
\begin{equation}
f(x,\phi) = e^{i \xi \phi}g(x)
\end{equation} 

And inserting it into \eqref{eqapradial1} we obtain 
\begin{equation}
\label{eqapradial1}
\left( \derivop{x^2}^2 + \frac{1}{x}\derivop{x} - \frac{\xi^2}{x^2} - 2\xi - x^2 + C \right) g(x) = 0 .
\end{equation}

We notice that 
\begin{equation}
\label{eqastuceradial}
\left(\derivop{x^2}^2 + \frac{1}{x}\derivop{x}\right) g(x) = \frac{1}{\sqrt{x}} \derivop{x^2}^2 \left(\sqrt{x}g(x)\right) + \frac{1}{4x^2}g(x),
\end{equation}
which once put into \eqref{eqapradial1} gives the following form 
\begin{equation}
\label{eqapradial2}
  \derivop{x^2}^2 \left(\sqrt{x}g(x)\right)  + \left( \frac{\frac{1}{4} - \xi^2}{x^2} - 2 \xi - x^2 + C \right)\left(\sqrt{x}g(x)\right) = 0.
\end{equation}

Here we recognize the differential equation giving generalized Laguerre functions given in \cite{olver_nist_2010}, table 18.8.1, 
\begin{equation}
 \left( \derivop{x^2}^2 +  \frac{\frac{1}{4} - \alpha^2}{x^2} - x^2 + 4s +2\alpha + 2 \right)h(x) = 0,
\end{equation}
where $\alpha$ is a real parameter stritly larger than $-1$, $s$ a positive integer and the solution $h$ is 
\begin{equation}
h(x) = e^{-\frac{x^2}{2}}x^{\alpha +1/2}L_{s}^{(\alpha)}(x^2).
\end{equation} 

Identifying $\alpha$ and $C$ in \eqref{eqapradial2} we find 
\begin{equation}
\alpha = \left\{ \begin{array}{rl}
 \pm & \abs{\xi} \text{ if } \abs{\xi} < 1 \\
 &\abs{\xi} \text{ if } \abs{\xi} \geq 1
\end{array} \right. ,
\end{equation}
and 
\begin{equation}
 C = \left\{ \begin{array}{c}
 4s + 2(\xi \pm \abs{\xi}) + 2 \text{ if } \abs{\xi} < 1 \\
 4s + 2(\xi + \abs{\xi}) + 2 \text{ if } \abs{\xi} \geq 1
\end{array} \right..
\end{equation}

Finally the solutions of equation \eqref{eqapradial1} are 
\begin{equation}
f(x, \phi) =  e^{i \xi \phi} e^{-\frac{x^2}{2}}x^{\alpha}L_{s}^{(\alpha)}(x^2).
\end{equation}

In the specific case of this paper we have $\xi = l_\perp$. Therefore we obtain 
\begin{equation}
\alpha = \left\{ \begin{array}{rl}
 \pm & \frac{1}{2}  \text{ if } l_\perp = \undemi \\
 &\abs{ l_\perp } \text{ otherwise } 
\end{array} \right.,
\end{equation}
and 
\begin{equation}
 C = \left\{ \begin{array}{cc}
 4s + 2 & \text{ if } l_\perp \leq 0 \\
 4(s + l_\perp) + 2 & \text{ if } l_\perp \geq 0
\end{array} \right..
\end{equation}


\section{Resolution of the Newtonian Hamilton-Jacobi problem of an electron in a constant uniform magnetic field} \label{apclassicalmotion}

\newcommand{\ur}{\vect{u}_r}
\newcommand{\uthet}{\vect{u}_\theta}
\newcommand{\dotthet}{\dot{\theta}}
\newcommand{\omc}{\omega_c} 
We work out the general solution, without assuming the center of motion, of the motion of an electron in a uniform constant magnetic field in polar coordinates using the Hamilton-Jacobi formalism of Newtonian mechanics. Although heavy, this way of obtaining a common result is interesting in view of the comparison with the quantum mechanical result, given the parenting between the Hamilton-Jacobi formalism and Hamiltonian quantum mechanics. 

Without loss of generality we restrict ourselves to a plane motion with polar coordinates $(r,\theta)$. The position vector $\vect{r}$ and the velocity $\dot{\vect{r}}$ are then expressed in the polar basis $(\ur, \uthet)$ by 
\begin{eqnarray}
\vect{r} & = & r\ur, \\
\dot{\vect{r}} & = & \dot{r}\ur + r\dot{\theta}\uthet.
\end{eqnarray}

The link with cartesian coordinates comes with $\ur = (\cos\theta,\sin\theta)$ and $\uthet = (-\sin\theta,\cos\theta)$. 

We choose to write the magnetic potential giving a field of flux intensity $B$ orthogonal to the plane of motion in a symmetric gauge with
\begin{equation}
\vect{A} = \undemi rB\uthet.
\end{equation}

Then the Lagrangian of an electron of charge $-e$ is given by 
\begin{equation}
L = \undemi m (\dot{r}^2 + (r\dot{\theta})^2 ) - \undemi er^2\dot{\theta}B.
\end{equation}

We readily see that $\theta$ is a cyclic coordinate as only its derivative participates in the Lagrangian. Therefore, its conjugate momentum is a constant of motion 
\begin{equation}
\label{eqpthet}
p_\theta= \pderiv{L}{\dotthet} = mr^2(\dotthet - \frac{\omc}{2}).
\end{equation}
Notice that $p_\theta$ is actually the angular momentum of the particle. 

We may now define the typical time scale $T$ and length scale $\lambda$ 
\begin{eqnarray}
(\omc/2)^{-1} & =& \left(\frac{eB}{2m}\right)^{-1},\\
\lambda & = & \sqrt{\frac{2p_\theta}{m\omc}},
\end{eqnarray}
which define an energy scale 
\begin{equation}
\epsilon = m(\lambda\omc/2)^2.
\end{equation}

We switch now to dimensionless coordinates 
\begin{eqnarray}
r \rightarrow  x = r/\lambda, \\
t \rightarrow \tau = \frac{\omc}{2}t,
\end{eqnarray}
and to a dimensionless Lagrangian 
\begin{equation}
 \tilde{L} = L/\epsilon = \undemi \left({x'}^2  + \theta'(1 + x^2)\right) - 1 - x^2,
\end{equation}
where $'$ denotes the derivation with respect to $\tau$ while $\dot{}$ was with respect to $t$. 

The momentum conjugated to $x$ is simply 
\begin{equation}
p_x = \pderiv{\tilde{L}}{x'} = x'.
\end{equation}

The Legendre transform of $\tilde{L}$ gives us the corresponding Hamiltonian 
\begin{equation}
\tilde{H} = p_x x' + p_\theta \theta' - \tilde{L} = \undemi\left(p_x^2 + \frac{1}{x^2} + 2 + x^2 \right),
\end{equation}
where we made obvious that $\theta$ was an ignorable coordinate by using \eqref{eqpthet} to get that $\theta' = 1/ x^2 + 1$ .

Let's now introduce the Hamilton characteristic function $W$ (see e.g. \cite{goldstein_classical_1980}), of which we consider only the $x$ dependence. We get the following Hamilton-Jacobi equation 
\begin{equation}
\left(\pderiv{W}{x}\right)^2 + \frac{1}{x^2} + 2 + x^2 = 2\tilde{E},
\end{equation}
\newcommand{\Etilde}{\tilde{E}}
where $\Etilde = E/ \epsilon$ is the dimensionless energy of the system. It follows that one gets the following Hamilton function 
\begin{equation}
W = \pm \int \dif{x} \sqrt{\Etilde - \frac{1}{x^2} - 2 - x^2}.
\end{equation}

Choosing $\Etilde$ as the new momentum we get that its conjugate coordinate is :
\begin{equation}
Q_{\Etilde} = \pderiv{W}{\Etilde} = \pm \int\frac{\dif{x}}{ 2\sqrt{\Etilde - \frac{1}{x^2} - 2 - x^2}}
\end{equation}
which integrates as
\begin{equation}
\label{eqQe}
 Q_{\Etilde} = \pm\undemi \arctan\left(\frac{a - x^2}{\sqrt{2 a x^2 - x^4 - 1}}\right)
\end{equation}
where $a = \Etilde -1$.

The denominator of the $\arctan$ argument in \eqref{eqQe} is necessarily positive since it is proportional to $ \left(\pderiv{W}{x}\right)^2$. By construction we now have $ Q_{\Etilde}' = \pderiv{\tilde{H}}{\Etilde} = 1 $. Integrating and equating to \eqref{eqQe} one obtains :
\begin{equation}
\label{eqx2}
\frac{a - x^2}{\sqrt{2 a x^2 - x^4 - 1}} = \pm \tan\left(2 \tau  + \theta_0\right)
\end{equation}

Solving for $x^2$ in \eqref{eqx2}, one obtains 
\begin{equation}
x^2 = a \pm \sqrt{a^2- 1 }\abs{\sin(2\tau + \theta_0)} .
\end{equation}

In order to keep a continuous trajectory, one will switch from the $+$ to the $-$ solution whenever the $\sin$ function switches as well. 

Finally, we use the conservation of angular momentum $\eqref{eqpthet}$ to obtain the equation for $\theta$ and switch back to the international unit system 
\begin{equation}
\label{eqsolution}
\begin{array}{l}
r = \lambda \sqrt{a + \sqrt{a^2- 1 } \sin \left(\omc t + \theta_0\right)}, \\
\theta =  \frac{\omc}{2}t + \arctan\left[ a\tan\left(\frac{\omc}{2}t + \frac{\theta_0}{2}\right) + \sqrt{a^2 -1}\right] + n \pi,
\end{array}
\end{equation}

with 
\begin{eqnarray}
\lambda & = & \sqrt{\frac{2p_\theta}{m\omc}}, \\
a & = & \frac{E}{\undemi p_\theta \omc} -1, \label{eqa}
\end{eqnarray} 
and 
\begin{equation}
n =  \frac{\pi}{2} + \text{floor}\left(\frac{\frac{\omc}{2}t + \frac{\theta_0}{2} - \frac{\pi}{2}}{\pi}\right).
\end{equation}

Thus, we obtained a solution depending only on the invariants of motion, $E$ and $p_\theta$, and a constant of integration $\theta_0$ depending on the initial conditions. 

Let's interpret the different trajectories. First, if we assume that $r$ or $\dot{\theta}$ is constant, from the expression of $p_\theta$ \eqref{eqpthet} we see that the other one is constant as well. From \eqref{eqsolution} we see that this is allowed only for $a =1 $ . In this case we find readily that 
\begin{eqnarray}
r & =& \lambda, \\
\theta & = & \omc t.
\end{eqnarray}

Using the relation between the conjugate momentum $p_\theta$ and the orthogonal "kinetic" momentum $p_\bot = mr\dot{\theta}$ 
\begin{equation}
p_\theta = r p_\bot - mr^2\frac{\omc}{2}.
\end{equation}

 Replacing $r$ by $\lambda$ in the above expression one obtains that the radius of the trajectory is the usual Larmor radius 
 \begin{equation}
 r = \lambda = \frac{p_\bot}{m\omc}
 \end{equation}
 
In this configuration the radial momentum is zero, and 
\begin{equation}
\label{eqEtheta}
E =  p_\theta\omc.
\end{equation}
In this particular configuration we see that $p_\bot$ is constant as well. 

We see that by construction : $a \geq 1$. Now, we consider the solutions for higher energies i.e. higher values of $a$. We see that they are all circles after computing the curvature radius using the formula 
\begin{equation}
\rho = \frac{(\deriv{r}{\theta}^2 + r^2)^{3/2}}{2\deriv{r}{\theta}^2 + r^2 - r\derivsec{r}{\theta}}.
\end{equation}

We obtain, after a lengthy but straightforward calculation, the general formula for the radius of curvature 
\begin{equation}
\label{eqrayontrajectoire}
\rho = \lambda\sqrt{\frac{a+1}{2}} = \sqrt{\frac{2E}{m\omc^2}}.
\end{equation}

Reminding that the magnetic field does not work, we find that the kinetic energy is constant and therefore these circles are circulated at constant speed $\omc\rho$. This allow us to find the center of each circle $\vect{r}_0$ by simple geometric considerations 
\begin{equation}
\label{eqpositioncentrenewt}
	\vect{r}_0 = \vect{r} - \frac{1}{\omc}\deriv{\vect{r}}{t}\wedge \frac{\vect{p_\theta}}{p_\theta},
\end{equation}
where $\vect{p_\theta}$ is the vectorial angular momentum, which here is orthogonal to the plane of motion. The distance of the center of motion to the origin of coordinates takes a simple form 
\begin{equation}
\norm{\vect{r}_0} = \sqrt{\rho^2 - \lambda^2} = \sqrt{2\frac{E - p_\theta\omc}{m\omc^2}}.
\end{equation}
 
Put differently, this gives a simple expression for the energy in terms of  
\begin{equation}
\label{eqnrjclassique}
E =  p_\theta\omc + \undemi m r_0^2 \omc^2.
\end{equation}

Thus we found the very intuitive result that, when the trajectory is centered on the origin of coordinates all the energy is stored in the angular momentum. However, we also see that the energy is "degenerate" with another invariant of motion, $r_0^2$, as in the quantum case. In any case, according to equation \eqref{eqrayontrajectoire} the radius of the trajectory is proportional to the square-root of the full energy.

\bibliography{Paper1-states_of_the_electron}

\end{document}